# Modélisation 1D-3Composantes de la réponse sismique d'une colonne de sol multicouche à comportement non linéaire.

## 1Directional-3Component seismic response modelling of a multilayer nonlinear soil profile.

M. P. Santisi d'Avila
*Université de Nice Sophia Antipolis, Laboratoire J. A. Dieudonné, Parc Valrose, 06108 Nice, France*

L. Lenti & J. F. Semblat
*Université Paris-Est, IFSTTAR, 58 Bd. Lefebvre 75015 Paris, France*

RÉSUMÉ : On propose une approche « 1D-3C » qui permet de modéliser la propagation unidirectionnelle (1D) simultanée des trois composantes (3C) d'un séisme, pour étudier la réponse sismique d'un sol horizontal multicouche, en utilisant une rhéologie du sol 3D non linéaire. Une loi de comportement élasto-plastique cyclique, de type Masing-Prandtl-Iwan, qui utilise uniquement la courbe de réduction du module de cisaillement pour caractériser le sol, a été implémentée dans un modèle aux éléments finis. La réponse sismique des profils de sol dépend de la polarisation du signal d'entrée, des propriétés élastiques et dynamiques du milieu et du contraste d'impédance entre les couches. La propagation d'un signal 3C induit un état multiaxial de contrainte et une interaction entre les composantes qui réduit la résistance du sol et augmente les effets non linéaires. Différents profils de sol dans la région de Tohoku (Japon) ont été soumis aux signaux sismiques enregistrés sur rocher affleurant ou en profondeur, pendant le séisme de Tohoku en 2011. La réponse sismique non linéaire de chaque profil de sol a été comparée dans les deux cas de superposition des effets de la propagation indépendante des trois composantes du signal (1D-1C) et 1D-3C.

ABSTRACT: We propose a one-directional three-component (1D-3C) approach to model the unidirectional (1D) propagation of a three component (3C) earthquake for seismic response analyses of horizontal multilayer soils, considering a 3D nonlinear constitutive behaviour for soils. An elasto-plastic cyclic constitutive behaviour of the Masing-Prandtl-Iwan type, using just the shear modulus decay curve for soil characterization, is implemented in a finite element scheme. Seismic response of soil profiles appears dependent on incident wave polarization, elastic and dynamic properties of medium and seismic impedance contrast between soil layers. Propagating a 3C signal induces a multiaxial stress interaction decreasing soil strength and increasing nonlinear effects. Soil profiles in the Tohoku area (Japan) are loaded by seismic signals recorded at outcrops or downhole, during the 2011 Tohoku earthquake. The nonlinear seismic response of each soil profile, represented in terms of acceleration, hysteresis loops and stress and strain profiles with depth, is compared in both cases of superposition of three one-component propagation effects (1D-1C) and 1D-3C.

MOTS-CLÉS: amplification des ondes sismiques, propagation des ondes en milieu non linéaire, mouvement fort.

## 1 INTRODUCTION

Les ondes sismiques qui se propagent dans les couches de sol superficielles, peuvent réduire la résistance du sol et augmenter les effets non linéaires. Le comportement non linéaire du sol peut amplifier ou réduire la réponse dynamique en surface, en fonction du taux de dissipation d'énergie.

Les modèles de propagation unidirectionnelle (1D) des ondes sont un moyen d'évaluer facilement le mouvement de terrain en surface, dans le cas de séismes forts. De nombreuses études ont été dédiées à la modélisation de la propagation unidirectionnelle des ondes de cisaillement pour des profils de sol multicouche, en considérant une seule composante du mouvement (polarisation 1C) et en utilisant comme rhéologie un modèle linéaire équivalent (Schnabel et al. 1972; Bardet et al. 2000), non linéaire pour sol sec (Bardet et al. 2001) ou non linéaire pour sol saturé (Lee et Finn 1978). Cela pour analyser les effets de site sismiques.

Le trajet de chargement tridimensionnel (3D) influence l'état de contrainte dans le sol et donc sa réponse sismique. Considérer un trajet de chargement 3D signifie représenter le comportement 3D non linéaire cyclique du sol, qui induit un couplage des trois composantes du mouvement, dont on est obligé de tenir compte (Li et al. 1992; Santisi d'Avila et al. 2012). Li (1992) a utilisé la loi de comportement 3D plastique cyclique de Wang, en termes de contraintes effectives, pour prendre en compte la pression interstitielle dans le sol dans un modèle 1D de propagation des ondes, aux éléments finis. Ce modèle rhéologique complexe nécessite un grand nombre de paramètres pour caractériser le sol.

Dans cette recherche, le comportement non linéaire du sol est représenté par un modèle de type Masing-Prandtl-Ishlinskii-Iwan (MPII), suggéré par Iwan (1967) et appliqué par Joyner (1975) et Joyner et Chen (1975) dans une formulation aux différences finies. L'implémentation de la loi de comportement non linéaire cyclique MPII dans un schéma de type éléments finis (code SWAP_3C) est présentée par Santisi d'Avila et al. (2012). La caractéristique principale de la formulation proposée est de résoudre le problème local déformation-contrainte en 3D, pour la simulation de la propagation unidirectionnelle des ondes sismiques dans le sol, en utilisant la rhéologie de type MPII, qui dépend seulement de propriétés mesurables en laboratoire. La fiabilité du modèle proposé, de propagation 1D-3C, a été évaluée en comparant les enregistrements à trois composantes du séisme de Tohoku en 2011 avec les signaux numériques (Santisi d'Avila et al. 2013).

Santisi d'Avila et al. (2012) ont analysé l'importance de la prise en compte les trois composantes (3C) du séisme. Ils ont analysé l'influence de différents paramètres sur la réponse du sol en utilisant des signaux incidents synthétiques. En analysant la propagation unidirectionnelle (1D) d'une onde incidente et en passant d'une à trois composantes, on observe, pour une déformation de cisaillement maximale fixée, une réduction du module de cisaillement et de la résistance du sol et une augmentation de la dissipation. La forme des boucles d'hystérésis ne varie pas à chaque cycle, pour un chargement à une composante, dans l'intervalle de déformation non linéaire stable. Dans le cas de chargement à trois composantes, pour une même gamme de déformations, la forme des boucles d'hystérésis change à chaque cycle. Les boucles d'hystérésis pour chaque direction horizontale sont altérées par l'interaction entre les composantes de chargement.

La différence principale entre la superposition des effets de trois mouvements de terrain à une composante (approche 1D-1C) et le modèle de propagation 1D-3C est analysée en termes d'histoire temporelle du mouvement, contrainte maximale et comportement hystérétique, avec plus de non linéarité et d'effets de couplage entre les composantes. Cet effet est plus évident pour petits rapports entre la vitesse des ondes de cisaillement et pression dans le sol et grands rapports entre le pic des composantes verticale et horizontale de l'onde incidente.

L'objectif de cette recherche est de confirmer, en utilisant des données réelles, les résultats de l'analyse paramétrique faite en utilisant des signaux synthétiques. La propagation simultanée de trois composantes du mouvement sismique (1D-3C) est comparée avec la superposition des effets de trois propagations d'une seule composante du mouvement (1D-1C), dans des profils de sol de la région de Tohoku (Japon), pour comprendre l'influence d'un trajet de chargement 3D et de la polarisation de l'onde incidente. On a utilisé des signaux sismiques ayant un rapport entre les pics en accélération verticale et horizontale supérieur à 70%, pour comprendre l'influence d'une composante verticale importante. L'influence des propriétés du sol et des ondes sismiques incidentes sur la réponse locale sont discutées.

## 2 MODÈLE DE PROPAGATION 1D-3C

Les trois composantes du mouvement sismique sont propagées dans un profil de sol, à comportement non linéaire, à partir de l'interface entre le bassin sédimentaire et le rocher sous-jacent, à comportement élastique. On suppose que le sol multicouche est infiniment étendu suivant les directions horizontales. Les ondes de cisaillement et de pression se propagent verticalement dans la direction $z$. Ces hypothèses conduisent à ne pas avoir de variation de déformation dans les directions $x$ et $y$. Le sol est supposé être un milieu continu et homogène, à chaque profondeur, en régime de petites déformations.

### 2.1 *Discrétisation spatiale*

La stratification de couches horizontales, parallèles au plan $xy$, est discrétisée en utilisant un schéma de type éléments finis (Fig. 1). On a adopté des éléments quadratiques linéiques à trois nœuds. En accord avec le modèle aux éléments finis, l'équation d'équilibre discrète s'écrit sous forme matricielle

$$\mathbf{M}\ddot{\mathbf{D}} + \mathbf{C}\dot{\mathbf{D}} + \mathbf{F}_{\text{int}} = \mathbf{F} \qquad (1)$$

où $\mathbf{M}$ est la matrice de masse, $\dot{\mathbf{D}}$ et $\ddot{\mathbf{D}}$ sont les vecteurs vitesse et accélération, respectivement, c'est-à-dire la première et la deuxième dérivée dans le temps du vecteur de déplacement $\mathbf{D}$. $\mathbf{F}_{\text{int}}$ est le vecteur des forces internes nodales. Le vecteur des charges $\mathbf{F}$ et la matrice de dissipation $\mathbf{C}$ sont non nuls quand on utilise une condition à la limite absorbante et ils sont nuls quand on impose le mouvement enregistré en profondeur comme condition à la limite.

Le problème différentiel d'équilibre (1) est résolu en accord avec les conditions de compatibilité et l'hypothèse de variation de déformation nulle dans les directions horizontales, avec une loi de comportement 3D non linéaire cyclique et avec les conditions aux limites décrites de suite.

La discrétisation de la colonne de sol en $n_e$ éléments quadratiques linéiques et donc en $n = 2n_e + 1$ nœuds (Fig. 1), qui ont trois degrés de liberté chacun, implique un vecteur de déplacement $\mathbf{D}$ de dimension $3n$, composé par trois blocs dont les termes sont les déplacements des $n$ nœuds dans les directions $x$, $y$ et $z$, respectivement. Les propriétés du milieu sont constantes dans chaque élément fini et dans chaque couche de sol. L'assemblage des matrices de dimension $(3n \times 3n)$ et des vecteurs de dimension $3n$ est fait de façon indépendante pour les trois sous-matrices de dimension $(n \times n)$ et pour les trois sous vecteurs de dimension $n$, respectivement, correspondant aux directions de mouvement $x$, $y$ et $z$.

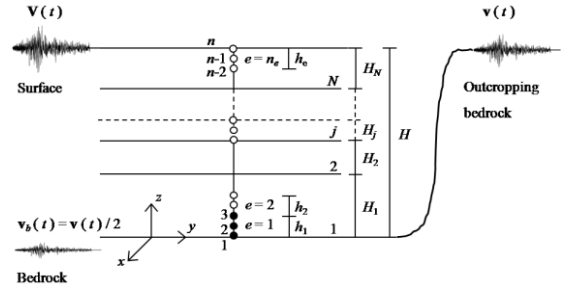

Figure 1. Discrétisation spatiale d'un bassin sédimentaire horizontal multicouche secoué à la base par un séisme à trois composantes, enregistré sur rocher affleurant et divisé par deux.

### 2.2 *Conditions aux limites*

Le système de couches de sol horizontales est limité en haut par la surface libre où on suppose nulles les contraintes normales. Le rocher à la base est modélisé comme un milieu élastique semi-infini. La condition à la limite suivante, appliquée à l'interface sol-rocher, permet à l'énergie d'être réfractée dans le milieu sous-jacent (Joyner et Chen 1975, Joyner 1975, Bardet et Tobita 2001), en tenant compte d'une rigidité finie du rocher:

$$-\mathbf{p}^T \boldsymbol{\sigma} = \mathbf{c}(\mathbf{v} - 2\mathbf{v}_b) \qquad (2)$$

Les contraintes normales à la base de la colonne de sol à l'interface sol-rocher sont $\mathbf{p}^T\boldsymbol{\sigma}$ et $\mathbf{c}$ est une matrice diagonale de dimension $(3 \times 3)$ dont les termes sont $\rho_b v_{sb}$, $\rho_b v_{sb}$ et $\rho_b v_{pb}$. Les paramètres $\rho_b$, $v_{sb}$ et $v_{pb}$ sont la densité du rocher et la vitesse des ondes de cisaillement et de compression dans le rocher, respectivement. Les trois termes des vecteurs $\mathbf{v}$ et $\mathbf{v}_b$ sont, respectivement, les vitesses inconnues à l'interface sol-rocher (nœud 1 dans la Fig. 1) et les données d'entrées en termes de vitesses des ondes dans le milieu élastique sous-jacent, dans les directions $x$, $y$ et $z$. L'onde incidente à trois composantes $\mathbf{v}_b$ peut être obtenue en divisant par deux les enregistrements sismiques au rocher affleurant (Fig. 1), en considérant l'effet de surface libre dans la roche supposée élastique linéaire.

Si on utilise les enregistrements en profondeur, contenant des ondes incidentes et réfléchies, dans ce cas le mouvement à l'interface sol-rocher (nœud 1 dans la Fig. 1) est connu et imposé comme condition à la limite.

### 2.3 *Modèle de comportement 3D non linéaire cyclique*

Le modèle rhéologique de Masing-Prandtl-Ishlinskii-Iwan adopté pour les sols a été sélectionné car il est tridimensionnel avec comportement non linéaire en charge et en décharge et surtout parce que le seul paramètre nécessaire pour caractériser le comportement hystérétique du sol est la courbe de réduction du module de cisaillement $G(\gamma)$ avec la déformation de cisaillement $\gamma$. C'est un mécanisme à plusieurs surfaces de plasticité pour sol sec, utilisé dans un intervalle de déformation de non linéarité stable. Les taux de déformation plus élevés ne sont pas bien représentés sans prendre en compte les conditions de sol non drainé.

### 2.4 *Discrétisation dans le temps*

Le modèle aux éléments finis et la non linéarité du sol demandent une discrétisation dans l'espace et dans le temps, pour permettre la résolution du problème. La loi de comportement incrémentale est linéarisée à chaque pas de temps. L'équation (1) est donc exprimée comme

$$\mathbf{M} \, \Delta \ddot{\mathbf{D}}_k^i + \mathbf{C} \, \Delta \dot{\mathbf{D}}_k^i + \mathbf{K}_k^i \, \Delta \mathbf{D}_k^i = \Delta \mathbf{F}_k \quad (3)$$

où l'indice $k$ indique le pas de temps $t_k$ et $i$ indique l'itération du processus de convergence à la solution. À chaque pas de temps $k$, l'équation (3) nécessite une résolution itérative, pour corriger la matrice de rigidité $\mathbf{K}_k^i$. Le processus de correction continue jusqu'à que la différence entre deux approximations successives se réduise à une tolérance fixée. Le pas suivant est ensuite analysé.

L'algorithme de Newmark, une approche implicite utilisée pour problèmes dynamiques, permet la résolution pas-à-pas (Newmark 1959; Hughes 1987). Les deux paramètres $\beta = 0.3025$ et $\gamma = 0.6$ assurent la stabilité inconditionnelle du processus d'intégration et une dissipation numérique qui atténue les modes à fréquence plus élevée (Hughes 1987).

## 3 COMPARAISON ENTRE 1D-3C ET 1D-1C

La réponse sismique d'un sol horizontal multicouche à la propagation d'un signal à trois composantes (approche 1D-3C) est comparée à la superposition des effets des trois composantes propagées de façon indépendante (approche 1D-1C).

Les enregistrements du mouvement du sol en surface pendant le séisme de Tohoku du 11 Mars 2011 (magnitude 9), et la stratigraphie des profils de sol sont fournies par le réseau accélérométrique K-Net. Les trois composantes du mouvement enregistré, Nord-Sud, Est-Ouest et Verticale, sont appelées respectivement $x$, $y$ et $z$ dans le modèle proposé.

La comparaison entre les enregistrements et les signaux numériques 1D-3C est montrée en termes de module de l'accélération en surface sur la Figure 2, pour le profil de sol FKS011 (Tableaux 1 et 2). Le mouvement incident est le signal enregistré sur rocher affleurant, c'est-à-dire à la surface du profil de type roche FKS015, ensuite divisé par deux (Tableau 1).

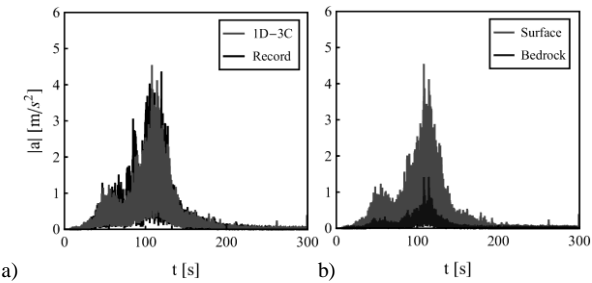

Figure 2. Module de l'accélération pendant le séisme de Tohoku : signal enregistré et numérique en surface (a); signal d'entré au rocher et signal numérique en surface (b), dans le cas FKS011/FKS015.

Tableau 1. Profil de sol et profil de type roche sélectionnés dans la région de Tohoku (Japan).

| Nom du site | Iwaky | Tanagura |
|---|---|---|
| Préfecture | Fukushimaken | Fukushimaken |
| Code du site | FKS011 | FKS015 |
| Distance épicentre (km) | 206 | 250 |
| Profondeur H (m) | 10.00 | 10.03 |
| $v_s$ moyenne (m/s) | 222 | 463 |
| min $\{v_p / v_s\}$ | 3.05 | - |

La comparaison entre les approches 1D-1C and 1D-3C est montrée en termes d'histoire temporelle en surface sur la Figure 3. Le signal incident en termes de module de l'accélération arrive amplifié à la surface du profil de sol analysé pour les deux approches 1D-1C et 1D-3C, mais le pic est réduit dans le cas 1D-3C et approxime mieux les enregistrements (Fig. 3).

Tableau 2. Stratigraphie et propriétés du sol pour le profil FS011.

| FKS011 | z (m) | $H_i$ (m) | $\rho$ (kg/m$^3$) | $v_s$ (m/s) | $v_p$ (m/s) | $\gamma_r$ (‰) |
|---|---|---|---|---|---|---|
| Sol sup. | 2.2 | 2.2 | 1430 | 100 | 700 | 0.800 |
| Limon | 3 | 0.8 | 1650 | 210 | 700 | 0.427 |
|  | 4 | 1 | 1720 | 210 | 1300 | 0.427 |
|  | 5.95 | 1.95 | 1660 | 330 | 1300 | 0.427 |
| Argile | 6.85 | 0.9 | 1810 | 330 | 1300 | 2.431 |
|  | 8 | 1.15 | 1970 | 330 | 1300 | 100 |
| Roche | 9 | 1 | 1980 | 590 | 1800 | 100 |
|  | 10 | 1 | 2060 | 590 | 1800 | 100 |

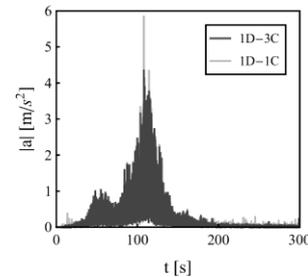

Figure 3. Module de l'accélération en surface pendant le séisme de Tohoku : résultats 1D-3C et 1D-1C pour le cas FKS011/FKS015.

La réponse locale du profil de sol FKS011 à un séisme à trois composantes est analysée par les profils avec la profondeur du module maximum d'accélération et de vitesse, de contraintes et déformations octaédriques et par boucles déformation-contrainte dans la couche plus déformée (Fig. 4). Les accélérations maximales (Tableau 3) et les vitesses sont plus élevées pour la combinaison de trois propagations 1C (approche 1D-1C). Les contraintes maximales sont réduites dans le cas 1D-3C et dans les couches plus souples les déformations maximales peuvent être plus élevées (Fig. 4).

Tableau 3. Accélérations enregistrées pendant le séisme de Tohoku en haut du profil de sol FKS011 et du profil de type roche FKS015 (signal d'entrée) et accélérations calculées en haut du profil de sol FKS011.

|  | Enregistrements | | 1D-3C | 1D-1C |
|---|---|---|---|---|
|  | Sol | Roche | Sol/Roche | |
| Profil | FKS011 | FKS015 | FKS011/FKS015 | |
| $a_x$ (m/s$^2$) | **3.74** | **1.36** | 3.78 | 3.97 |
| $a_y$ (m/s$^2$) | 3.12 | 1.01 | 3.92 | 4.33 |
| $a_z$ (m/s$^2$) | 3.00 | 0.58 | 1.64 | 0.89 |
| $|a|$ (m/s$^2$) | **4.47** | **1.42** | 4.55 | 5.72 |

La courbe déformation-contrainte de cisaillement de premier chargement (Fig. 4) est obtenue, dans le cas de propagation 1C, par la courbe de réduction du module de cisaillement. Quand les déformations de cisaillement sont plus élevées que la limite élastique, on observe, sous chargement cyclique, des boucles dans le plan déformation-contrainte, avec hystérésis. La réponse cyclique du sol en termes de contrainte et déformation de cisaillement dans la direction $x$, dans les deux cas de signal d'entrée à trois composantes (1D-3C) et de propagation de la seule composante $x$ du signal (1D-1C) est comparée sur la Figure 4. En passant d'une à trois composantes, pour une déformation de cisaillement maximale fixée, le module de cisaillement diminue et la dissipation augmente. La résistance du matériau est inférieure dans le cas de chargement 3D plutôt que dans le cas de cisaillement simple, représenté par la courbe de premier chargement. Modéliser la propagation 1D d'un séisme à trois composantes permet de prendre en compte les interactions entre les composantes de cisaillement et de pression de la charge sismique. Un état de contrainte triaxial en milieu non linéaire, dû à un chargement 3D cyclique, induit des effets de couplage multiaxial.

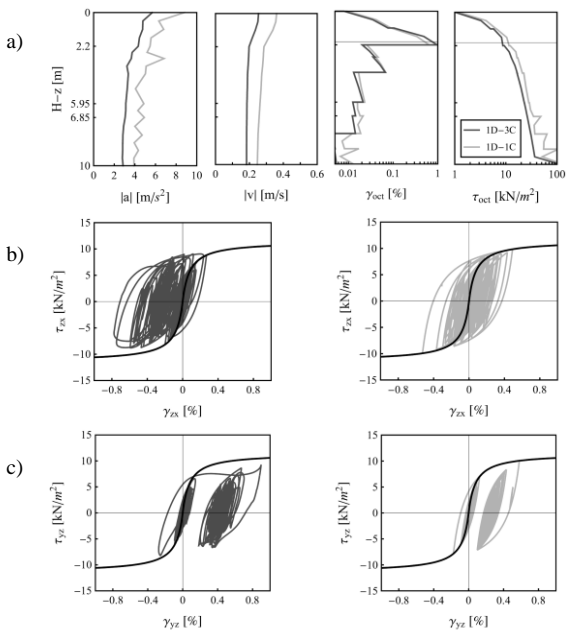

Figure 4. Réponse sismique 1D-3C et 1D-1C pendant le séisme de Tohoku, dans le cas FKS011/FKS015: profils avec la profondeur du module maximum d'accélération et vitesse, de déformation et contrainte octaédriques (a); boucle d'hystérésis à 2 m de profondeur pour les directions de cisaillement *x* (b) et *y* (c).

La forme des boucles d'hystérésis ne change pas à chaque cycle, sous chargement 1C, dans un intervalle de déformation de non linéarité stable. Sous chargement 3C, la forme des boucles d'hystérésis change à chaque cycle, même dans l'intervalle de non linéarité stable sous chargement 1C. Les boucles d'hystérésis pour chaque direction horizontale sont altérées du fait de l'interaction entre les composantes de chargement. Ce résultat confirme l'analyse faite par Santisi d'Avila et al. (2012) en utilisant des signaux synthétiques.

La différence plus évidente entre les approches 1D-1C et 1D-3C est observée dans l'histoire temporelle du mouvement en surface, contrainte maximale et comportement hystérétique, avec plus de non linéarité et d'effets de couplage entre les composantes.

## 4 CONCLUSIONS

On propose un modèle géomécanique de propagation des trois composantes des ondes sismiques pour des profils de sol 1D (approche 1D-3C). Ce modèle permet l'évaluation des effets locaux dans le cas de mouvement fort.

La loi de comportement 3D de type Masing-Prandtl-Ishlinskii-Iwan (MPII) est implémentée dans un schéma de type éléments finis, qui modélise un bassin sédimentaire horizontal multicouche, sous chargement cyclique. Cette rhéologie MPII a été adoptée pour le sol car elle est tridimensionnelle, non linéaire soit en charge qu'en décharge et parce que la courbe de réduction du module de cisaillement permet à elle seule d'identifier le comportement hystérétique du matériau.

La combinaison de trois analyses 1D-1C indépendantes est comparée à l'approche 1D-3C proposée, pour des profils de sol dans la région de Tohoku (Japan), secoués par le séisme de Tohoku en 2011. Les effets d'amplification à la surface sont moins importants dans le cas 1D-3C à cause des non linéarités et du couplage tridimensionnel du mouvement. L'état de sollicitation multiaxiale induit une réduction de résistance du matériau et une dissipation plus importante. La forme des boucles d'hystérésis change à chaque cycle dans le cas 1D-3C, même pour des déformations dans une gamme de non linéarité stable sous chargement 1C.

Les effets de la polarisation du mouvement incident et du trajet de chargement 3D sont observables avec l'approche 1D-3C, qui permet l'évaluation de paramètres non mesurés de mouvement, contrainte et déformation, au long du profil de sol, pour mieux identifier les effets non linéaires. Le coefficient de Poisson influence les propriétés de dissipation du sol et donc la réponse sismique locale. La polarisation des ondes sismiques incidentes, en particulier le rapport entre l'amplitude maximale des composantes verticale et horizontale, affecte le taux de dissipation d'énergie et les effets d'amplification. En particulier, un faible rapport entre la vitesse des ondes de cisaillement et pression dans le sol, et donc un coefficient de Poisson bas, et un rapport élevé entre le pic des composantes verticale et horizontale augmentent l'interaction mécanique triaxiale et changent progressivement la taille et la forme des boucles d'hystérésis à chaque cycle. Les déformations maximales sont observées à l'interface entre les couches, où, le long du profil de sol, les ondes rencontrent de forts contrastes d'impédance. Les effets non linéaires sont plus évidents dans la direction horizontale où le mouvement est plus faible. Elle est donc la plus influencée par le couplage 3D du mouvement.

L'extension du modèle 1D-3C proposé, pour permettre d'atteindre des déformations plus élevées, sera le prochain défi permettant l'analyse des effets non linéaires du sol saturé.

## 5 REMERCIEMENTS

## 6 RÉFÉRENCES